# THE EFFICIENCY OF HARMONIC EMISSIONS EXCITED BY ENERGETIC ELECTRONS IN CORONAL LOOPS

Mehdi Yousefzadeh,[1, *] Alexey Kuznetsov,[2] Yao Chen,[1, 3] and Mahboub Hosseinpour[4]

[1]*Institute of Frontier and Interdisciplinary Science, Shandong University, Qingdao, Shandong 266237, People's Republic of China*
[2]*Institute of Solar-Terrestrial Physics, Irkutsk 664033, Russia*
[3]*School of Space Science and Physics, Shandong University, Weihai, Shandong 264209, People's Republic of China*
[4]*Faculty of Physics, University of Tabriz, Tabriz, Iran*

## ABSTRACT

Magnetic reconnection is a key process that drives the energy release in solar flares. This process can occur at multiple locations along the coronal loop. The reconnection generates energetic electrons capable of exciting wave modes and emissions as they propagate through the loop. In this follow-up study, we investigate the influence of the injection site location of these energetic electrons—either at the looptop (LT) or at the leg of the loop around a footpoint (FP)—on the excitation of wave modes especially the second harmonic emissions (X2) in coronal loops. Our simulations reveal that the injection location significantly impacts the spatial distribution and intensity of excited wave modes. When electrons are injected at the LT, electromagnetic X2, and Z modes dominate along the loop, with minimal excitation of Langmuir waves (Yousefzadeh et al. 2021; 2022). Conversely, the present study reveals that injection close to FP leads to a strong Langmuir wave excitation throughout the loop, particularly as electrons ascend toward the LT. We

Corresponding author: Mehdi Yousefzadeh
yousefzadeh@sdu.edu.cn

find that X2 and Z modes are consistently excited at the injection site with different intensities, regardless of the injection location. However, electron injection near the FP scenario creates favorable conditions for significant Langmuir wave generation, potentially leading to plasma emission under specific circumstances. These findings emphasize the importance of electron injection location in determining the properties of the excited and emitted waves in solar coronal loops.

* E-mail: m.yousefzadeh6@gmail.com

1. INTRODUCTION

Solar radio bursts are primarily generated by coherent emission mechanisms, where plasma instabilities amplify specific wave modes, producing intense electromagnetic radiation. One of the prominent radiation mechanisms responsible for direct amplification of electromagnetic waves in solar and astrophysical contexts is the Electron Cyclotron Maser Emission (ECME), initially proposed by Twiss (1958). This mechanism has proven highly relevant in a variety of astrophysical scenarios, including solar flares, Earth's auroral kilometric radiation (AKR), and solar spike bursts (Wu 1985; Treumann 2006).

The ECME arises from the resonant interaction between energetic electrons and the cyclotron motion in a strong magnetic field. However, at the fundamental cyclotron frequency ($\Omega_{ce}$) the emitted waves often experience strong gyromagnetic absorption, inhibiting their escape. This limitation is reduced for harmonics of higher orders (e.g., $s \geq 2$), where the absorption coefficient decreases significantly, allowing the radiation to propagate and potentially escape (Melrose & Dulk 1982). Furthermore, as highlighted by Wu & Lee (1979), ECME emission is predicted to be most efficient when the wave frequency is close to the local electron cyclotron frequency, under the condition that $\omega_{pe}/\Omega_{ce} < 1$. The preferential polarization of emitted waves is the right-handed X-mode, as opposed to the O-mode which stems from the rotational consistency of the wave-particle interaction (Sharma & Vlahos 1984).

Another crucial mechanism for radio emission involves the generation of Langmuir waves, which often serve as an intermediary in producing plasma radiation. Langmuir waves are longitudinal electrostatic oscillations driven by energetic electron beams propagating through a plasma. These waves are commonly generated through beam-plasma instabilities when a population of energetic electrons with a velocity distribution exceeding the thermal electron speed interacts with the background plasma (Chen et al. 2022; Zhang et al. 2022; Yousefzadeh 2025). In the context of solar and stellar radio bursts, these Langmuir waves can undergo nonlinear processes, such as wave-wave coupling or scattering by ion sound waves, to produce electromagnetic radiation at the fundamental and harmonic frequencies of the plasma frequency ($\omega_{pe}$). This plasma emission mechanism is distinct from ECME, as it relies on the interplay between beam-plasma instabilities and the natural oscillatory modes of the plasma, rather than the direct coupling of cyclotron motion with electromagnetic waves. ECME and plasma emission represent two of the most studied coher-ent mechanisms in astrophysical plasmas, offering complementary insights into the physics of high-energy electron populations and the conditions under which radio bursts are produced. The widely accepted model for plasma emission suggests, that a beam of energetic electrons leads to a kinetic bump-on-tail instability, which can be explained through a multistage nonlinear process initially proposed by Ginzburg & Zhelezniakov (1958). The plasma emission mechanism (involving beam-excited Langmuir waves) is known to be responsible for the generation of solar type III radio bursts (Melrose 2017).

To explore the excitation of harmonic emissions, Yousefzadeh et al. (2021) proposed a comprehensive three-step numerical scheme. The first step involves using the nonlinear force-free field (NLFFF) extrapolation method (Wheatland et al. 2000; Wiegelmann 2004; Wiegelmann et al. 2006) which provides a detailed magnetic field structure in a large-scale topology of an active region. In the second step, the guiding center approximation (GCA; Northrop 1963) is applied to track the evolution of millions of electrons along the magnetic loop. This method reveals electrons' spatial and temporal velocity distribution functions (VDFs) within a large-scale loop. The third step employs the particle-in-cell (PIC) method to simulate the kinetic instabilities caused by the obtained VDFs.

This multistep numerical scheme represents a foundational approach for investigating the multiscale processes involved in solar radio bursts. By integrating the macroscopic dynamics of electron transport within large-scale coronal loop structures with the microscopic physics of kinetic instabilities, this framework provides a pathway to understand the mechanisms that drive radio wave emissions. It offers insights into how energetic electrons propagate and interact with the magnetic topology, ultimately leading to the excitation of coherent radiation in solar and astrophysical plasmas.

In earlier work by Yousefzadeh et al. (2021), the focus was primarily on analyzing the VDFs of electrons at the loop-top (LT) at a particular moment, specifically regarding the variations in scattering intensities. Their study revealed that the electron VDFs at the LT exhibited distinct strip-like along with loss-cone features, which served as the primary sources of kinetic instabilities. Their results highlighted the efficient excitation of the X2 mode through ECME, a mechanism critical for resolving the escaping difficulty of such emissions. However, this analysis was limited to the LT region, leaving the potential spatial and temporal variations of the emission properties along other parts of the loop unexplored.

To expand on this, Yousefzadeh et al. (2022) investigated emissions' spatial and temporal evolution across different loop sections. Their study revealed that the X2 mode was predominantly excited in the upper parts of the loop with the most intense emission occurring at the topmost region. In contrast, the Z mode exhibited significant excitation across all loop regions, whereas the fundamental X mode remained undetected.

Building on this prior research, the present study investigates a new scenario in which the injection location of energetic electrons is shifted from the LT to a region along the leg of the loop, closer to a footpoint (FP). This approach aims to explore how the changes in injection location affect the VDFs, kinetic instabilities, and the resulting emission properties, providing a more comprehensive understanding of the mechanisms that drive emissions within coronal loops. This is the main objective of the current study.

## 2. THE THREE-STEP NUMERICAL SCHEME

The numerical methodology employed in this study builds upon the three-step scheme introduced in previous works (Yousefzadeh et al. 2021; 2022), with distinct adjustments to address the current research focus. The selected active region and loop were chosen to establish a realistic magnetic environment for the simulation. The loop represents a typical field line in the corona above the pre-eruption flux rope structure. While a more complex twisted configuration might be interesting, a simple loop configuration can reveal the essential physics effectively. The large-scale magnetic topology of the selected untwisted coronal loop was derived using the NLFFF extrapolation method, as detailed in earlier studies. The plasma density distribution along the loop was inferred from the hydrostatic model established in Yousefzadeh et al. (2022), ensuring consistency with the magnetic field distribution and corresponding plasma frequency ratio $\omega_{pe}/\Omega_{ce}$. For the present study, two specific sections along the loop —section A ($\omega_{pe}/\Omega_{ce} = 0.4$) and section B ($\omega_{pe}/\Omega_{ce} = 0.16$)— were selected for the detailed analysis. Notably, the ratio of $\omega_{pe}/\Omega_{ce}$ in section A is 2.5 times greater than in section B, providing contrasting environments for evaluating kinetic instabilities and emission processes. Figure 1(a) depicts the Helioseismic and Magnetic Imager (HMI; Schou et al. 2012) data overplotted with the result of extrapolation, while Figure 1(b) highlights the spatial location of the two sections along the selected loop.

We note that the processes of magnetic reconnection in flares and the mechanism of the particle acceleration are beyond the scope of this study, i.e., we focus on the propagation of the energetic particles along a selected magnetic flux tube; for that purpose, the magnetic field structure produced by the NLFFF extrapolation is appropriate. In this work, the injection location of energetic electrons is shifted from the LT, as studied previously, to the leg of the loop near the FPs within section B. The electrons were impulsively injected and followed using the GCA method ([Northrop 1963](#)). This approach tracks the transport and evolution of 4 million electrons, initially distributed according to a Maxwellian function with a thermal velocity of 0.24c (corresponding to the energy of ∼15.6 keV). VDFs were obtained for sections A and B at a specific time post-injection to examine the influence of the injection location on the development of kinetic instabilities. The pitch angle scattering process was simulated by randomly modifying individual electrons' pitch angles at intervals corresponding to a scattering time scale ($\tau$). Following [Yousefzadeh et al. (2022)](#), $\tau$ was set to ∼2s, representing the intermediate scattering scenario, consistent with observational constraints on the behavior of ∼15 keV electrons during large solar flares ([Chen & Petrosian 2013](#)).

PIC simulations were performed using the VPIC code ([Bowers et al. 2008a](#); [2008b](#); [2009](#)) to study the kinetic instabilities induced by the obtained VDFs in each section. The background electron-proton plasma was modeled as a Maxwellian distribution (∼2 MK), with the density ratio of energetic to background electrons fixed at $n_e/n_0 = 0.05$. The background magnetic field was aligned with the $z$-axis ($B_0\hat{e}z$), and the wave vector ($\vec{k}$) was placed in the $xOz$ plane. The simulation domain was configured as [512, 512] $\Delta$, with a cell size $\Delta \sim 2.7\lambda_D$ (where $\lambda_D$ is the Debye length of the background electrons). The simulation duration extended to 1000 $\omega_{pe}^{-1}$ for cases that reached saturation. In each cell, 1000 macroparticles per species were employed, maintaining charge neutrality throughout the simulation.

3. RESULTS OF THE DYNAMICS OF ENERGETIC ELECTRONS ALONG THE LOOP

Electrons were injected in the lower part of the loop, close to the FP location, as shown by the red square labeled "I" in Figure 1(b). Following their injection, electrons began to move bidirectionally along the loop. The VDFs were analyzed for two selected regions of the coronal loop: section A and section B (see Figure 1(b)). These VDFs were obtained by plotting the phase-space positions of all electrons within the selected areas. Figure 2(a)-(b) shows a snapshot of these evolved VDF along with the Maxwellian distribution of background electrons overlaid as the central dark-red region. Once the injection location was considered near the FP, significant electron losses occurred on one side of the loop during the early stages of the simulation. In contrast, on the opposite side, electrons travel further along the loop appearing in other sections after a certain time. Notably, in addition to the initial strip interacting with the background particles, two distinct strip-like features appeared in the VDFs for section B. At the same time, a hollow beam-like structure was observed in section A (see Figure 2(a)-(b)).

The strip-like features in section B reflect electrons undergoing bidirectional motion, originating from the bouncing behavior of energetic electrons within the loop. Only electrons within specific pitch angle ranges (satisfying the mirroring condition) were reflected back and forth within the loop. The strong magnetic field in the lower parts of the loop affects the visibility of a full strip in section A. The location of electron injection near the footpoint, where the magnetic field is stronger, makes it more challenging for electrons with low parallel energy to ascend to higher altitudes where the field strength is weaker. Therefore, only electrons with the highest parallel velocity are observable in the other strips of section A, shaping a hollow beam-like feature. The general morphology of the VDFs is taking longer to develop its main features compared to Yousefzadeh et al. (2021). The selected VDFs were obtained at 1.4s postinjection time which is slightly longer than our earlier studies. Similar strip features were also observed in earlier research by White et al. (1983), which examined a single-dimensional magnetic field line and analyzed analytical expressions of VDFs.

Here, the PIC simulations reveal that different wave modes are excited in distinct regions of the coronal loop, influenced by the initial VDF and local plasma configuration characteristics. Figure 3 illustrates the temporal evolution of energy changes for the six electromagnetic field components alongside the change in electron kinetic energy, starting from its initial value, with a negative sign, i.e. $-\Delta E_k(t) = -(E_k(t) - E_k(0))$. The negative sign is needed for plotting convenience to show an increase in $-\Delta E_k(t)$, while electromagnetic field components also increase. The Gaussian filter method can evaluate energy curves for each mode (Yousefzadeh et al. 2021; 2022; 2025). It identifies the $\omega$ and k ranges from the dispersion diagram plot. A Gaussian profile is then assumed, and an inverse Fourier transform determines the energy distribution in time and space. Finally, the energy is integrated spatially to create the mode's temporal energy profile (e.g. see Figure 3). Figure 4 depicts the wave distribution in the $\vec{k}(k_\parallel, k_\perp)$ space, highlighting the dominant wave intensities for specific wave vectors. The movie further shows the temporal evolution of wave distribution.

Figure 5 presents the $\omega-k$ dispersion analysis across different propagation angles where $\theta$ is the angle between $\vec{k}$ and $\vec{B}_0$. Here, we discuss these simulation outputs for different sections.

### 3.1. *The Excitation and propagation of the Waves at LT Location (Section A)*

Starting with a Maxwellian electron distribution at the leg of the loop close to the FP, the injection process results in a hollow beam-like distribution in section A. This distribution appears compact in the $v_\parallel$-$v_\perp$ space, making the plasma instability driven by $\partial f/\partial v_\parallel > 0$, the dominant mechanism where electrons shifting primarily along the parallel velocity ($v_\parallel$; see Figure 2(c) and the accompanying movie).

The energy field components are shown in Figure 3(a) along the total kinetic energy of all electrons ($-\Delta E_k$). The temporal energy profile of the beam-Langmuir (BL) wave mode is overplotted with a dashed line and its energy level reaches the maximum value of $\sim 5 \times 10^{-3} E_{k_0}$. The simulation can be divided into two stages based on the temporal evolution of energy in the field components. In the first stage (0 to 75 $\omega_{pe}^{-1}$), the $E_z$ component undergoes significant growth, marking the early phase of wave excitation. At its peak, the energy associated with $E_z$ constitutes a few percent of the total kinetic energy of all electrons. Alongside $E_z$, other field components also experience rapid increases in energy, growing by several orders of magnitude during this phase. An exception is the perpendicular E component ($E_y$), which shows a different behavior, increasing at a slow but steady rate throughout the entire simulation. In the second stage (75 $\omega_{pe}^{-1}$ till the end of simulation), all field components transition into a post-saturation phase, where their energy contributions stabilize. A steady and uniform behavior across the field components characterizes this stage. No significant energy fluctuations occur, indicating wave-particle interactions have reached a quasi-steady state.

Langmuir waves in section A propagate primarily in the beam direction, spreading across a broad $k$-space. Some waves reflect in the negative $k_\parallel$ direction (backward Langmuir waves), but these are less intense compared to their forward-moving counterparts (see Figure 4(a)). Figure 5(a) focuses on excited wave modes in Section A and includes the analytical dispersion curves for the magnetoionic X and Z modes, along with the Langmuir wave. This analysis reveals a substantial increase in wave activity mainly in $E_z$ direction. Figure 5(a) shows the dominant excited mode is the BL waves. At the same time, there are almost no indications of Z or harmonic modes in the perpendicular $E_y$ direction.

3.2. *The Excitation and propagation of the Waves at around FP Location (Section B)*

The velocity distribution becomes more diffuse, forming a strip-like structure in the $v_\parallel$–$v_\perp$ space in section B. Here, the condition of $\partial f/\partial v_\perp > 0$ is satisfied across a wider velocity range, with $v_\perp$ spanning from 0.2c to ~0.4c, compared to around ~0.15c in section A. Additionally, the strip-like feature covers a wider range from 0 to over 0.2c in the parallel velocity direction, while section A remains around ~0.2c. In this region, electrons shift primarily in the perpendicular velocity direction ($v_\perp$; see Figure 2(d) and the accompanying movie).

The resonance curve condition for the ECM plays a fundamental role in understanding the energy transfer mechanisms responsible for wave excitation in plasma environments. This condition establishes the relationship between the wave frequency, electron motion, and the local magnetic field, allowing efficient interaction between the waves and electrons. Specifically, it identifies regions in electron velocity space where the local cyclotron harmonic frequency closely matches the wave frequency. Electrons in these regions most probably resonate with the wave, leading to significant emission and absorption coefficients essential for ECM instability and emission processes. In Figure 2(b), we examine the resonance curves for the Z and X2 modes using the VDFs obtained from section B of the simulation. The analysis reveals that the strip-like features in the VDF are the primary drivers of the excitation for these modes. These strip-like features reveal sharp gradients in the distribution function (i.e. here $\partial f/\partial v_\perp > 0$), serving as key drivers of plasma instabilities. This is supported by the resonance curve condition of the excited wave modes (Figure 2(b)) and the perpendicular evolution of the derived VDF (Figure 2(d)). Notably, the first strip has no significant impact, as it passes through the central region of the background distribution, failing to interact effectively with the waves. The other strips align with the resonance curve, providing the appropriate conditions for efficient energy transfer and wave excitation. This highlights the importance of specific VDF structures in shaping the system's ECM emissions characteristics.

Figure 3(b) shows the energy evolution for the six field components under conditions where multiple wave modes, including BL, the second harmonic (X2), and Z-mode waves, contribute to the energy dynamics. During the initial phase (0 to 250 $\omega_{pe}^{-1}$), the $E_y$ component shows rapid energy growth a few orders of magnitude higher than the other components. Other field components are growing but in lower magnitude compared to the $E_y$. This simultaneous increase in the energy of electric and magnetic components highlights the interaction of multiple wave modes with the electron population. All three modes can be seen in the distribution of the maximum intensity of the field components in the $k_\parallel$–$k_\perp$ space (see Figure 4(b)). The $\omega$–$k$ dispersion analysis result is shown in Figure 5(b).

Transverse Z-mode waves are the dominant feature in this region, reaching a peak energy level of approximately $\sim 6 \times 10^{-3} E_{k_0}$, although Langmuir waves are also present. Consequently, the growth rate of the Z-mode in section B exceeds that of the Langmuir waves, in which the latter achieves a peak energy of around $10^{-4} E_{k_0}$. The X2 mode is excited in a quasi-perpendicular direction, attaining a maximum intensity of about $\sim 2 \times 10^{-5} E_{k_0}$. Due to the exponential nature of wave growth, even a small difference in growth rates can lead to significant changes in energy changes between the different wave modes. When Figures 4 and 5 are considered together, they provide a comprehensive view of the characteristics of the excited modes. The data in section A clearly show no evidence of perpendicular Z modes or harmonic modes within the system. However, a strong presence of Langmuir waves is observed, particularly in the $E_z$ direction. These differences between the two sections are influenced by the plasma-to-cyclotron frequency ratio and, more significantly, by the distinct nonthermal electron velocity distributions in these regions.

## 4. SUMMARY AND DISCUSSION

Our findings demonstrate that the location of energetic electron injection within a coronal loop plays a crucial role in determining the excitation of wave modes and emission properties. X2 and Z modes are consistently excited at the injection site, which typically corresponds to the reconnection region. However, these modes' spatial distribution and intensity vary significantly based on the injection location. For LT injection, electromagnetic X2 and Z modes dominate throughout the loop, with substantial excitation across nearly all regions (Yousefzadeh et al. 2021; 2022). In contrast, when electrons are injected near the FP, the excitation of X2 and Z modes is largely restricted to the injection region, with minimal activity detected at higher loop altitudes.

A comparison of these scenarios reveals notable differences in emission dynamics. In the LT injection case, ECME (e.g., X2 and Z modes) dominate, while Langmuir waves are weakly excited along the loop. This suppression of Langmuir waves is likely due to the transition of electrons from regions with higher frequency ratios to those with lower values. This shift suppresses the conditions necessary for Langmuir wave generation. Conversely, the reflected electrons tend to form a beam-like distribution in FP injection, leading to a significant generation of Langmuir waves throughout the loop. This especially becomes more evident as the electrons move upward from the FP toward the LT, with the strongest activity occurring at the highest altitude. During this journey, the frequency ratio increases from approximately 0.16 to 0.4, improving conditions for Langmuir wave generation and enhancing the possibility of plasma emission. While this satisfies the primary conditions for plasma emission, the consistently low-frequency ratio throughout the loop limits their development. These results underscore the critical influence of the injection site on the mechanisms of wave excitation and emission within coronal loops.

By observing the VDFs produced by the GCA code, it is clear that the general morphology of the VDFs remains stable for about ~0.3 seconds. This indicates that the EVDF is not a transient distribution of electrons but persists over longer time scales. This stability enhances its value for further analysis, meaning the distribution remains in the environment long enough to generate waves potentially. Tests with lower-energy electrons (~7.6 keV, about half of the case presented) reveal similar large-scale morphologies of the EVDF, though with lower average energy and a longer time required to reach a steady state. This suggests that within a specific energy range, the modes of wave excitation remain consistent, even though their intensity varies. Note that, the incoherent emission from hot thermal electrons is present and detectable; however, it is significantly weaker than both the coherent emission and the incoherent emission from higher-energy electrons (hundreds of keV to several MeV). Therefore, the parameters and assumptions of our model are consistent with known observational data.

We further investigated the same FP injection scenario with higher density ratios of energetic electrons to the background, resulting in a slight increase in the intensity of the X2 mode at the injection site. However, this increase remains relatively weak compared to LT injection scenarios. As a result, the possibility of escaping radio waves due to efficient harmonic emission of ECME is significantly lower when the injection occurs in the leg of the loop around FP, compared to the LT injection site.

The results of our simulations suggest that shifting the electron injection site from the looptop to the footpoint might significantly alter the dominant emission mechanism. LT injection leads to direct X2-mode generation via ECME, resulting in strong X-mode polarization and emission at twice the local cyclotron frequency. In contrast, FP injection favors plasma emission through Langmuir wave generation, which primarily exhibits O-mode polarization at the plasma frequency or its harmonic. These differences not only affect emission intensity but also influence polarization and directivity, offering potential observational diagnostics for distinguishing between the two cases.

Given these distinctions, modern radio instruments capable of high-resolution spectral and polarization measurements, such as Expanded Owens Valley Solar Array (EOVSA), and Siberian Radio Heliograph (SRH), provide promising possibilities for testing these predictions. However, the current lack of sufficient spatial resolution makes direct observational verification challenging. Future observational studies, combined with independent measurements of coronal plasma parameters, could determine if specific reconnection sites along flaring loops preferentially excite particular waves, offering new insights into flare energy release and particle acceleration processes.

## ACKNOWLEDGMENTS

The research results are sponsored by the China/Shandong University International Postdoctoral Exchange Program and the Youth Specialty Category grant under project number ZR2024QD206. The authors acknowledge the open-source VPIC code by Los Alamos National Labs (LANL) and the Super Cloud Computing Center (BSCC, URL: http://www.blsc.cn/) for providing computational resources. The work of A.K. was supported by the Ministry of Science and Higher Education of the Russian Federation. M.Y. acknowledges Dr. David Tsiklauri (SU) for his comments and discussion.

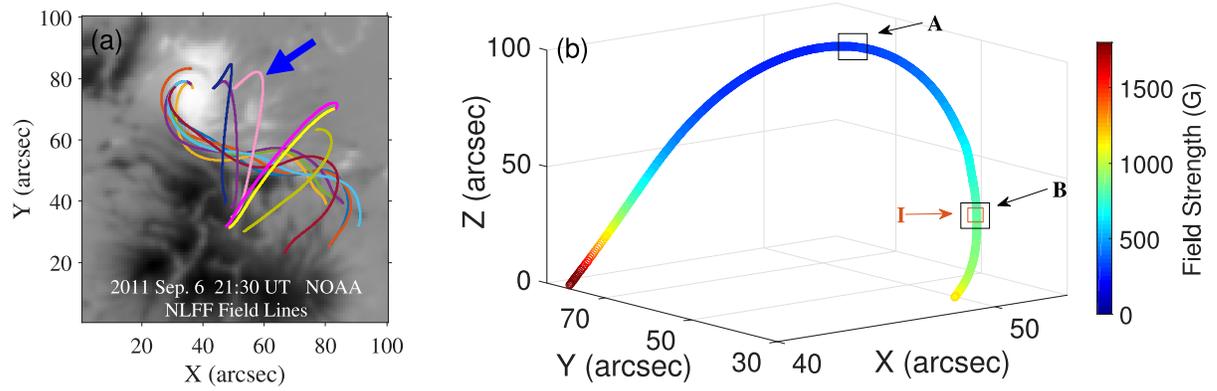

**Figure 1.** (a) The magnetogram image of HMI for AR 11283 is overplotted with various magnetic field lines. A blue arrow indicates the selected loop. (b) The magnetic field strength of the selected field line. The sections denoted by letters A and B are the areas where the VDFs will be analyzed, and the letter I marks the injection region.

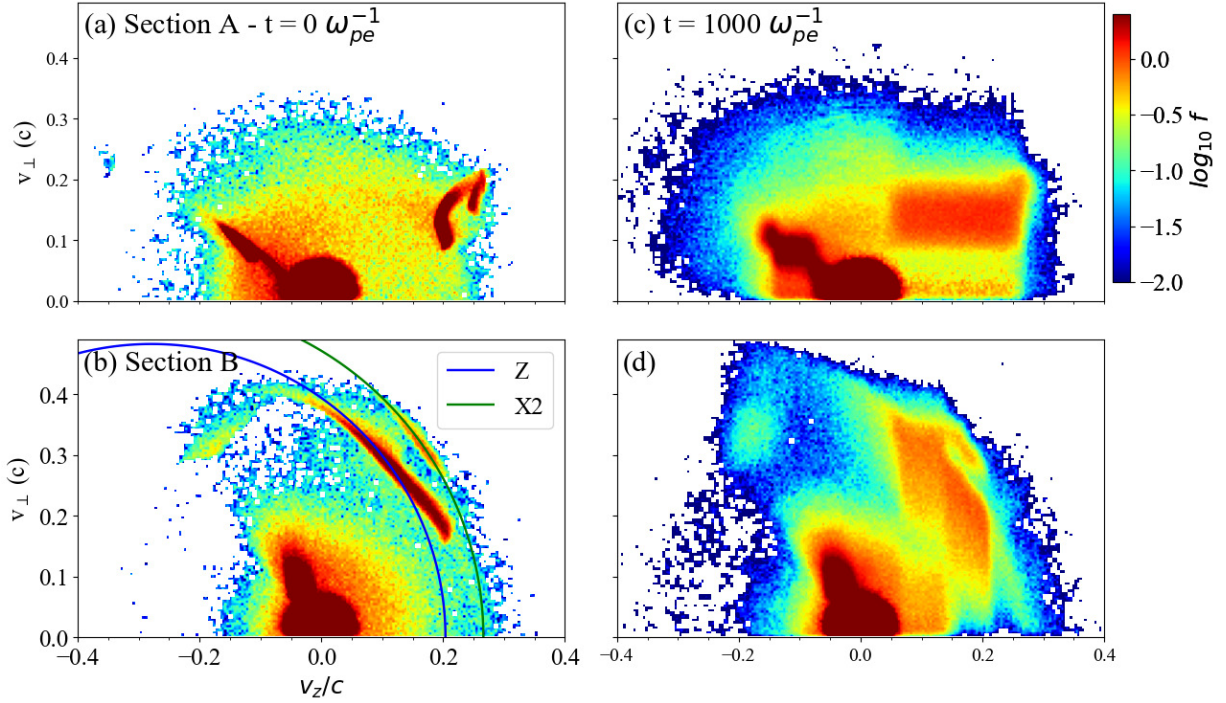

**Figure 2.** (a)-(b) The obtained VDFs at a specific time (1.4s post-injection), within the selected sections of the loop overplotted with the distribution of background particles. The resonance curves corresponding to the X2 and Z modes are shown in panel (b). The parameters used to plot these curves are as follows: for X2, the wave frequency $\omega = 1.85\ \Omega_{ce}$, the total wave number $k = -1.98\ \Omega_{ce}/c$, the propagation angle $\theta = 75°$, and the harmonic number $n = 2$; and for Z they are $0.95\ \Omega_{ce}$, $-1\ \Omega_{ce}/c$, $75°$, and 1, respectively. The top two panels (a) and (c) are for section A at two different times $t = 0$ and $1000\ \omega_{pe}^{-1}$, while the bottom two panels (b) and (d) are for section B at the same times. A video of this figure starts at time $t = 0\ \omega_{pe}^{-1}$ and extends until the simulation ends at $t = 1000\ \omega_{pe}^{-1}$. The real-time duration of the video is ∼10s.

(An animation of this figure is available.)

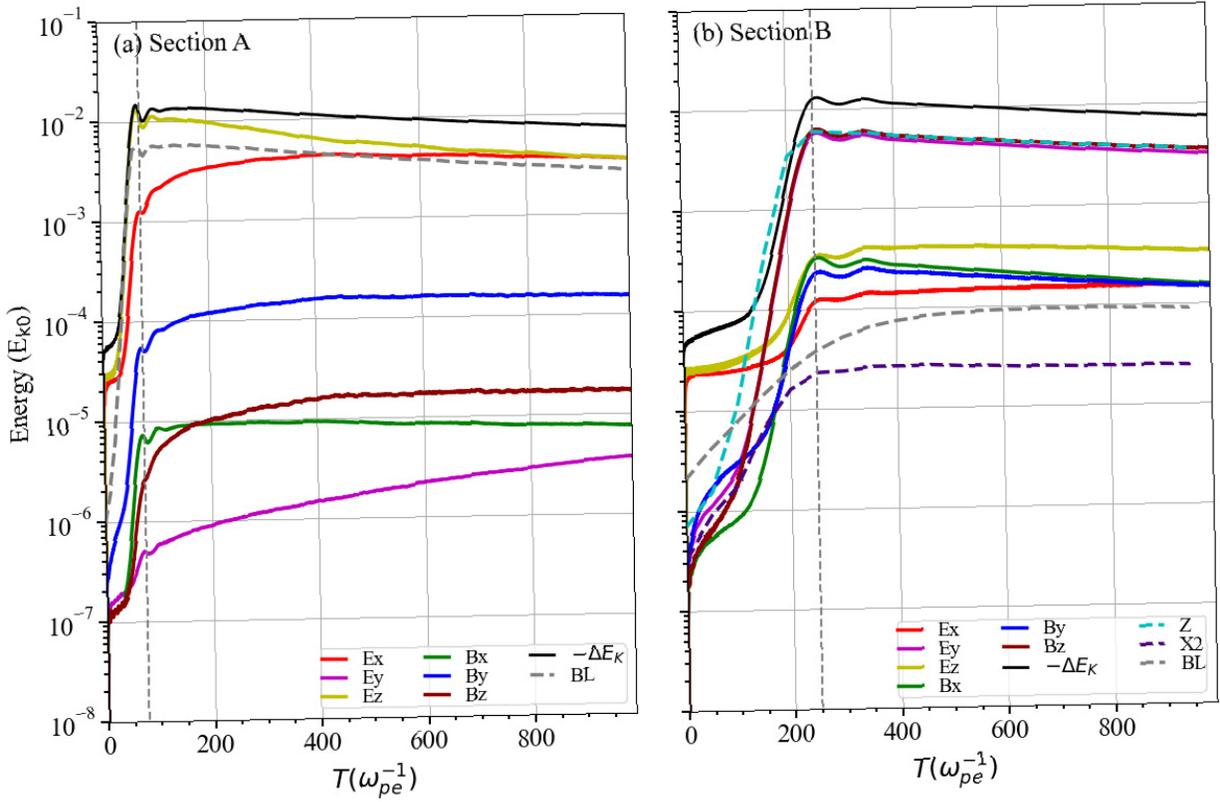

**Figure 3.** (a)-(b) illustrate the temporal evolution of energy changes for the six field components, alongside the negative variation in total electron energy ($-\Delta E_k$) for sections A and B, respectively. Additionally, the temporal profiles of energies for various wave modes (BL, Z, and X2), normalized to the total energy of energetic electrons ($E_{k0}$) are overplotted. The vertical line indicates the times at which the different field components reach their maximum energy level.

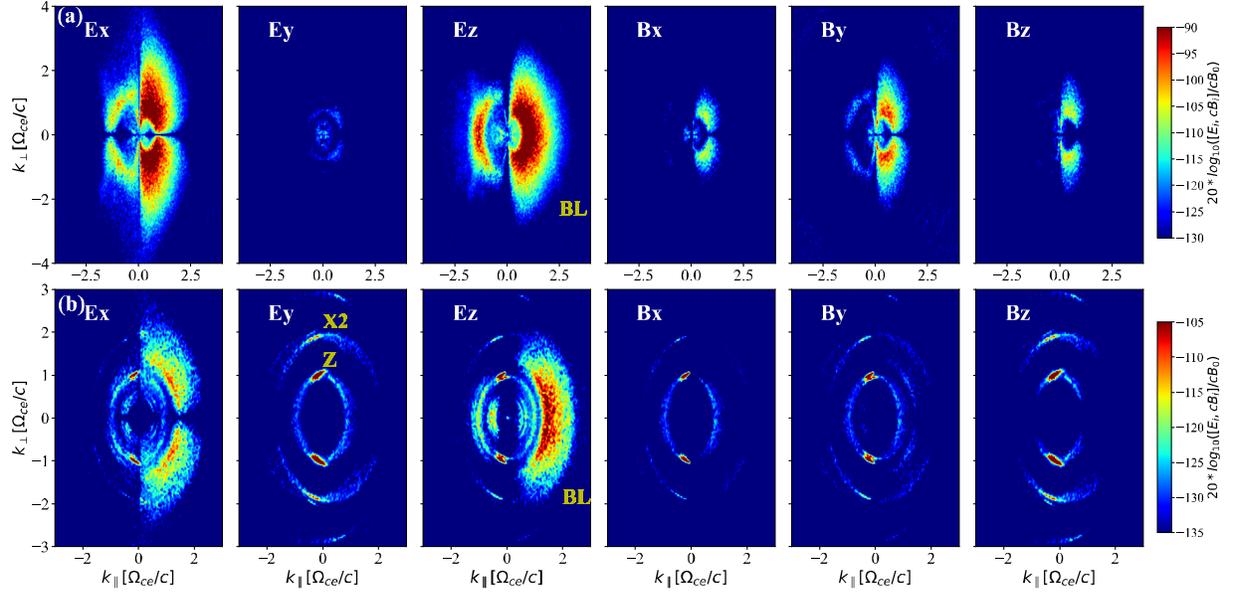

**Figure 4.** Upper Panels: The intensity maps of the six field components in the $k_\parallel$-$k_\perp$ space for section A; Lower panels: Correspondingly, these panels present the distribution of the maximum intensity of the field components for section B. A video of this figure starts at time t = 0 $\omega_{pe}^{-1}$ and extends until the simulation ends at t = 1000 $\omega_{pe}^{-1}$. The real-time duration of the video is 10s.
(An animation of this figure is available.)

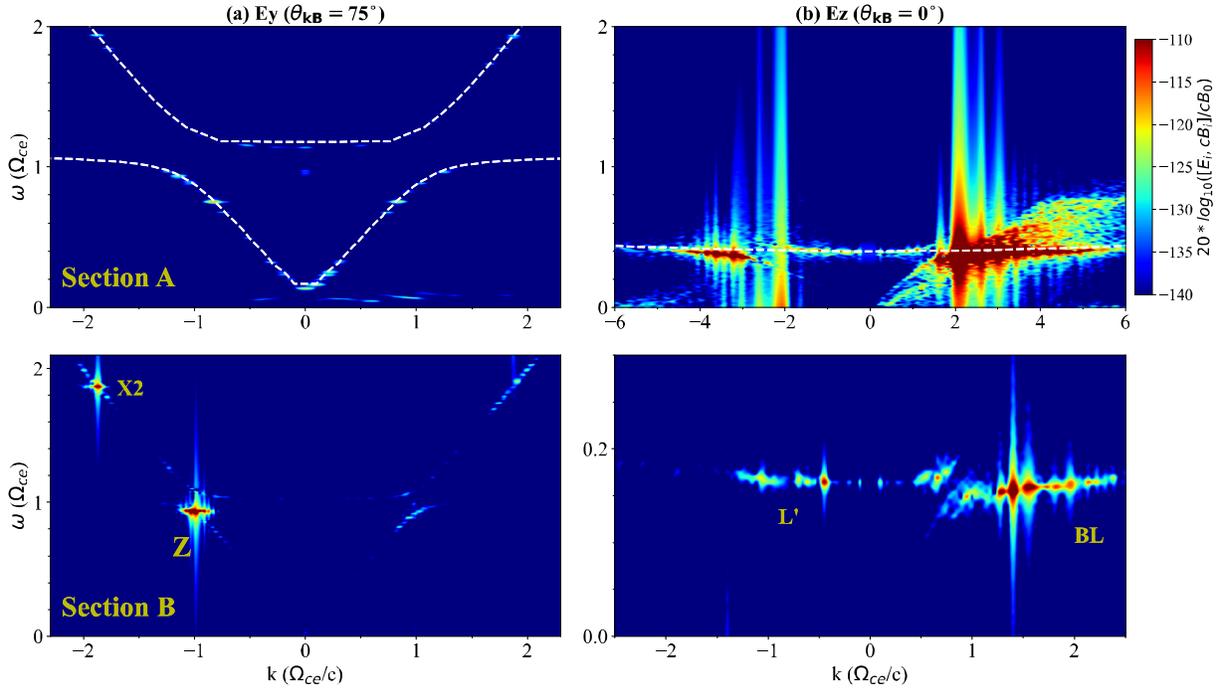

**Figure 5.** The upper panels show the dispersion diagrams for the field components: (a) the perpendicular electric field (Ey) and (b) the parallel electric field (Ez) for section A. The lower panels present the dispersion plots of these components for section B, highlighting the presence of BL, L′, X2, and Z modes. The dashed lines represent dispersion curves based on magnetoionic theory.